\documentclass[letterpaper,american,reprint,amsmath,amssymb,prd,showpacs]{revtex4-1}
\usepackage[latin9]{inputenc}
\usepackage{enumitem}
\usepackage{amsmath}
\usepackage{esint}

\makeatletter

\pdfpageheight\paperheight
\pdfpagewidth\paperwidth

\DeclareFontEncoding{LGR}{}{}
\DeclareRobustCommand{\greektext}{%
  \fontencoding{LGR}\selectfont\def\encodingdefault{LGR}}
\DeclareRobustCommand{\textgreek}[1]{\leavevmode{\greektext #1}}
\ProvideTextCommand{\~}{LGR}[1]{\char126#1}

\providecommand{\tabularnewline}{\\}

\usepackage{dcolumn}
\usepackage{bm}

\makeatother

\usepackage{babel}
\begin{document}

\title{Cosmology from conservation of global energy}

\author{Herman Telkamp}

\address{Jan van Beverwijckstraat 104, 5017 JA Tilburg, The Netherlands}
\email{herman\_telkamp@hotmail.com}

\begin{abstract}
\noindent It is argued that many of the problems and ambiguities of
standard cosmology derive from a single one: violation of conservation
of energy in the standard paradigm. Standard cosmology satisfies conservation
of local energy, however disregards the inherent global aspect of
energy. We therefore explore conservation of the quasi-local Misner-Sharp
energy within the causal horizon, which, as we argue, is necessarily
an apparent horizon. Misner-Sharp energy assumes the presence of arbitrary
mass-energy. Its conservation, however, yields ``empty'' de Sitter
(open, flat, closed) as single cosmological solution, where Misner-Sharp
total energy acts as cosmological constant and where the source of
curvature energy is unidentified. It is argued that de Sitter is only
apparently empty of matter. That is, total matter energy scales as
curvature energy in open de Sitter, which causes evolution of the
cosmic potential and induces gravitational time dilation. Curvature
of time accounts completely for the extrinsic curvature, i.e., renders
open de Sitter spatially flat. This explains the well known, surprising,
spatial flatness of Misner-Sharp energy, even if extrinsic curvature
is non-zero. The general relativistic derivation from Misner-Sharp
energy is confirmed by a Machian equation of recessional and peculiar
energy, which explicitly assumes the presence of matter. This relational
model enhances interpretation. Time-dilated open de Sitter is spatially
flat, dynamically close to $\Lambda$CDM, and is shown to be without
the conceptual problems of concordance cosmology.
\end{abstract}
\maketitle

\section{Introduction}

Conservation of energy in the expanding universe is controversial.
It is incompatible with basic elements of standard cosmology, in particular
the non-uniform scaling of matter densities and the non-stationarity
of FLRW spacetimes. In addition, there is no unanimously agreed upon
notion of gravitational energy within general relativity, which of
course affects the subject of energy conservation too. Fundamentally
this is due to the non-local nature of gravitational energy, which
has no coordinate independent local representation like the stress-energy
tensor. Standard cosmology is derived from geometry (the Friedmann
equation) and conservation of local energy (the continuity equation).
The global aspect of energy is in fact ignored. This in itself may
be totally justified, considering the homogeneity of the Universe
on the large scale. The issue, though, is that conservation of global
energy likely constrains the evolution of density parameters, while
the continuity equation permits different forms and mixtures of matter,
potentially giving rise to odd behavior, like, e.g., violation of
energy conservation, the coexistence of multiple ``causal'' horizons
and inconsistency of the different forms of redshift. It is quite
curious that the continuity equation ensures ongoing conservation
of local energy of radiation, while within an infinitesimal time interval
the same radiation fails conservation of energy due to redshift, a
specifically non-local feature. This indeed suggests that matter in
global perspective cannot have arbitrary equation of state, and that
the scaling must relate to global properties of spacetime. 

The other issue is non-stationarity. Since FLRW spacetimes are non-stationary,
it is often argued that energy can not be conserved in the expanding
Universe. While this indeed applies to most universe models, it is
not generally the case. As Florides showed \citep{Florides}, there
are exactly six FLRW spacetimes (the ones with constant curvature)
which have a static representation, namely: Minkowski, Milne, de Sitter
(with $k=-1,0,1$, for open, flat and closed geometry, respectively),
and anti-de Sitter. The common Friedmann form of these spacetimes
is given by Florides as 
\begin{equation}
\dot{a}^{2}+k+Ka^{2}=0,\label{Florides}
\end{equation}
where $K=-\Lambda/3$ represents the cosmological constant. Of these
six spacetimes, the flat de Sitter universe is cosmologically highly
relevant in the early (inflating) and late universe, and total vacuum
energy within the de Sitter event horizon is indeed conserved. So
conservation of global energy is not totally alien to cosmology. The
issue of course is that all six spacetimes of Eq.(\ref{Florides})
are ``empty'' of matter. 

So there are certainly arguments within standard cosmology to give
up on conservation of energy in the Universe. Violation of conservation
of energy is however not the single fundamental problem of standard
cosmology. As we argue, multiple ambiguities and inconsistencies arise
from violation of conservation of global energy. Indeed it turns out
that imposing conservation of global energy resolves other well known
issues too. Moreover, this dramatically constrains the solution space
of FLRW spacetimes, which in turn gives rise to reconsideration of
standard notions of energy. 

The purpose of this note is to explore the cosmological implications
of conservation of global energy within the context of general relativity.
This requires a notion of non-local energy within the causal horizon.
We approach this in two parallel ways: through the Misner-Sharp quasi-local
energy \citep{MisnerSharp} and through a Machian expression of total
energy \citep{TelkampPhysRevD.94.043520}. The former becomes global
if applied to the causal sphere of the Universe, while the latter
considers the causal sphere by principle. The two lead to consistent
energy equations of total energy. However, the meaning of energy terms
in the Misner-Sharp equation is rather implicit, while the relational
Machian expression explicitly models recessional and peculiar energy
of matter within the causal horizon. The Misner Sharp equation, on
the other hand, enables full treatment of the subject within general
relativity. The Machian notions mainly serve interpretation. In addition
to conservation of energy, we make as little as possible assumptions,
e.g., just the presence ($\rho_{m}>0$) of otherwise unspecified matter.
Initially, we will be concerned exclusively with physical consistency
of the model. Then, we will evaluate the results for observational
and theoretical relevance. 

In terms of local energy, the causal horizon may seem a rather irrelevant
feature of the Universe; the horizon does not appear in the standard
formulation of the Friedmann equation. But evidently, the distance
to this horizon is a key parameter of global energy. Identification
of the causal horizon is therefore an essential step. This is not
quite a trivial matter, since the notion of causal horizon is ambiguous
in standard cosmology. This ambiguity however dissolves in the de
Sitter case. Not surprisingly, conservation of total energy indeed
implies a de Sitter universe (open, flat, or closed). 

The most significant aspect of this result, however, is that we assume
the presence of matter, yet obtain ``empty'' de Sitter solutions.
This may seem a departure from general relativity, but it follows
immediately from application of the apparent horizon to the Misner-Sharp
energy, which holds in any spherically symmetric spacetime. 

Another main result is that the curved solutions, open and closed
de Sitter, show evolution of the gravitational potential, therefore
necessarily induce gravitational time dilation. This prompts adjustment
of the (constant) metric coefficient $g_{tt}\rightarrow g_{t't'}(a)$.
As it turns out, curvature of time accounts completely for extrinsic
curvature in open and closed de Sitter. This means that, taking gravitational
time dilation into account, all three de Sitter solutions are necessarily
spatially flat. Cosmologically, flat de Sitter, $k=0$, can be identified
with the spacetime of the freely falling comoving observer, while
open de Sitter, as we show, matches accelerated observational coordinates
on the past light cone of the observer. In this way ``time dilated
open de Sitter'' can match CMB observations (due to flat space) as
well as SNIa observations (due to gravitational time dilation). Open
de Sitter has a present deceleration $q_{0}=-\frac{1}{2}$, which
makes it quite close in expansion dynamics to \textgreek{L}CDM in
the redshift range of SN1a observations. Open de Sitter actually outperforms
\textgreek{L}CDM in BIC score on the Union2.1 compilation.

Open de Sitter has many favorable properties and leads to natural
explanations of both unidentified ``dark'' matter and vacuum energy.
It may thus provide a viable cosmological model, without the conceptual
problems of \textgreek{L}CDM.

\section{Concordance cosmology}

The \textgreek{L}CDM model, even while observationally favorable,
is physically not understood for multiple reasons, i.e., the horizon
problem, coincidence problem, flatness problem, the age problem and
the unidentified dark matter and vacuum energy. Inflation theory successfully
addresses some of these problems, but does not simplify cosmology
and relies on the hypothetical inflaton field, which is not physically
understood either. One may add that there are several ambiguities
and fundamental issues in the theoretical concepts and assumptions
underlying standard cosmology. This regards of course in particular
violation of the conservation of energy in the standard paradigm,
but also a number of less recognized issues, addressed as follows.

In standard cosmology, matter densities scale differently due to different
equation of state, i.e., $\rho_{dust}\propto a^{-3}$ and $\rho_{radiation}\propto a^{-4}$.
The extra factor $a$ of radiation accounts for redshift due to expansion
of space. Any difference in dilution rates, however, not only sacrifices
conservation of energy, it also violates the equivalence of the different
forms of mass-energy, at least in the strict interpretation due to
Eddington, where the different forms are just different appearances
of one and the same property, i.e., energy \citep{Eddington,StanfEncyclEquivalencePrinciple}.
One would therefore expect dust and radiation to scale identically
to maintain mass-energy equivalence. Then, there is the question of
coordinates being physical or not. Cosmological time is commonly assumed
to represent physical (clock) time of a comoving observer. Not only
at present, but through all era of the universe. The implicit assumption
here is constancy of the potential, thus the absence of gravitational
time dilation (the lapse function in the FLRW-metric being set to
$g_{tt}=1$). But is this assumption justified? Then there are a number
of ambiguous concepts involved, like the unclear relationship between
sources and curvature, the coexistence of different horizons with
``causal properties'' \citep{Ellis:2015wdi}, and seemingly incompatible
forms of redshift \citep{Chodorowski:2009cb,Faraoni:2009buREDSHIFT,Melia:2012ic}.
At the bottom of this, there is the question of the representation
of energy, i.e., local versus non-local \citep{Szabados2009}. Within
general relativity the non-localness of gravitational energy is recognized,
however not satisfactorily resolved. 

\smallskip{}

One may not subscribe to all of these concerns, but the list is simply
too long to not question certain fundamentals of standard cosmology.
On the other hand, there may well be only few underlying causes of
the issues jointly. Violation of conservation of energy is likely
one, since it is connected with all other issues. Thus there is reason
to reconsider notions of energy, in particular the scaling of the
energy densities of matter. We will however reverse order by considering
conservation of global energy, and make inferences therefrom. This
involves identification of the causal horizon.

\section{Causal horizon\phantomsection\label{Section:Causal Hor}}

An observer's causal horizon is the spherical boundary between particles
which are connected with the observer, and the particles which are
not \citep{Ellis:2015wdi}. The question is when exactly a particle
is connected? For our purposes, we will narrow this to the question
which masses contribute momentarily to the gravitational potential
at the origin. Rather than trying to answer this immediately, we will
note the ambiguities surrounding the subject and point out three consistent
cases.

The cosmological event, particle and apparent horizon all three have
certain ``causal properties''. The ambiguity arises from the limited
causal relevance of these horizons individually, as well as from their
coexistence at different distances (for instance, the apparent horizon
and the event horizon in accelerating universes). The issue centers
around the temporal aspect of connection. One could argue that the
particle horizon, the farthest distance we receive particles from
\textit{today}, is a clear demarcation of causality \citep{Ellis:2015wdi}.
However, in a presently accelerating universe, the massive source
of these particles has long disappeared behind the event horizon,
so is disconnected forever and does not contribute anymore to the
potential. By definition of the apparent horizon \citep{Rindler},
an inward directed photon emitted at present time from a galaxy crossing
the observer's apparent horizon has momentarily zero physical speed,
as measured in the frame of the observer; nothing from this galaxy
is moving into our direction \textit{now}. Thus at this point the
mass-energy of this galaxy, or from anything beyond, does not contribute
to the gravitational potential in the frame of the observer, so is
disconnected. This is actually what one expects to happen at the causal
boundary \textit{now}. But in case of the apparent horizon this is
only momentarily so; it is not a null surface in general. An inward
directed photon, emitted from a galaxy crossing this horizon, will
still move towards the observer in terms of the comoving speed, thus
will eventually reach the observer nevertheless. In the case of accelerating
universes, it is only beyond the farther located event horizon that
photons, emitted now, will not reach the observer ever. The event
horizon therefore unambiguously demarcates the forever disconnected
region. Even so, a photon emitted exactly from the event horizon now
will only reach the observer in infinite time. So one may question
the immediate relevance of this photon \citep{Ellis:2015wdi}, and
in fact of all particles located in between apparent and event horizon,
as these do not contribute at present to the potential in the observer's
frame. So in fact, all three horizons have causal characteristics,
but none is unequivocally a boundary of causal connection on its own.

The interpretative differences between the horizons however disappear
in a few particular cases \citep{Rindler}. In de Sitter the apparent
horizon coincides with the event horizon (the particle horizon diverges).
In this case, \textit{disconnected now} unambiguously means \textit{disconnected
forever}. Similarly, in a radiation universe the apparent horizon
coincides with the particle horizon (the event horizon diverges).
In that case, \textit{connected now} means \textit{connected forever}.
This universe however does not match solution set Eq.(\ref{Florides}),
which is surprising in a way. It is worthwhile mentioning that there
is one other consistent case: the static Minkowski universe, where
all three horizons diverge. Minkowski is part of the solution set
of Eq.(\ref{Florides}).

Note that the apparent horizon, least of all considered causal, is
the single horizon being part of all consistent cases. It is always
nearest and always exists \citep{FaraoniPhysRevD.84.024003}. Most
importantly, it is the only horizon with an instant character, which
is an indispensable feature to meaningfully act as a causal boundary
now. It therefore seems reasonable to assume that the causal horizon
necessarily is an apparent horizon. The reverse of course need not
be the case, since we expect only specific solutions of the apparent
horizon to meet conservation of global energy.

\section{Global energy equation}

\subsection{Misner-Sharp quasi-local energy}

Representation of the non-local gravitational energy is an issue in
general relativity, which primarily is a local theory. It is well
known that the gravitational field vanishes in the inertial frame,
along with local spatial curvature. Indeed, gravitational energy is
intimately connected with curvature, but a single general definition
of gravitational energy is lacking. To overcome this, various notions
of quasi-local mass have been introduced for different spacetimes.
The Misner-Sharp mass $E_{MS}(R)$ applies to all spherically symmetric
spacetimes and expresses the \textit{total internal energy} contained
in a sphere of radius $R$ \citep{MisnerSharp}. It is the sum of
internal kinetic energy and internal potential energy. Note that in
the cosmological context both recessional motion and peculiar motion
contribute to total internal energy $E_{MS}$. The requirement of
spherical symmetry fits FLRW universes, which are homogeneous and
isotropic, but still strictly spherical by the presence of a gravitational
causal horizon at proper distance $R_{g}$. In this way, the Misner-Sharp
energy $E_{MS}(R_{g})$ provides a notion of global energy within
the radius of the Universe. 

As motivated above, we tentatively consider the causal horizon to
be an apparent horizon. We make no other assumptions, just that the
Universe contains matter $\rho_{m}>0$ (massive particles, radiation,...),
which implies non-zero energy of both recessional and peculiar motion.
The areal radius $R_{g}\equiv ar_{g}$ of the apparent horizon is
(in units where $c=1$) 
\begin{equation}
R_{g}^{2}=\frac{1}{H^{2}+\kappa a^{-2}}=\frac{1}{H^{2}-\frac{8}{3}\pi G\rho_{k}},\label{RA-1}
\end{equation}
where $H=\sqrt{\frac{8}{3}\pi G\rho}$ is the Hubble parameter, with
total density $\rho$, and where $\kappa=k/R_{g0}^{2}$ is the Gaussian
curvature, with index $k=-1,0,1$ for spatially open, flat and closed
geometries, respectively \citep{Rindler}. Note that curvature energy
density $\rho_{k}$ is always removed from the denominator of Eq.(\ref{RA-1}).
Misner-Sharp energy includes all internal energy and is defined \citep{MisnerSharp}
\begin{equation}
E_{MS}(R)=\frac{R}{2G}(1-\nabla^{a}R\nabla_{a}R).\label{E MS}
\end{equation}
A well known, surprising property of Misner-Sharp energy is \citep{FaraoniPhysRevD.84.024003,Binetruy:2014ela}
\begin{equation}
E_{MS}(R)=\frac{4}{3}\pi R^{3}(\rho-\rho_{k}),\label{E MS-1}
\end{equation}
i.e., the \textit{r.h.s.} shows a flat volume, even if $\rho_{k}\neq0$.
In case of the apparent horizon, $\nabla^{a}R_{g}\nabla_{a}R_{g}=0$
\citep{FaraoniPhysRevD.84.024003}. Misner-Sharp energy within the
apparent horizon $E_{g}\equiv E_{MS}(R_{g})$ therefore reduces to
(the Schwarzschild mass)
\begin{equation}
E_{g}=\frac{R_{g}}{2G}>0.\label{EAH}
\end{equation}
From this simple result alone it follows that if total energy $E_{g}$
is conserved, then 
\begin{equation}
R_{g}=const,\label{Rg=00003Dconst}
\end{equation}
which unambiguously points at the de Sitter universe, where the constant
Misner-Sharp total energy density acts as cosmological constant. This
solution resolves the ambiguity between apparent and event horizon
in accelerating universes, as the two coincide. De Sitter ($k=-1,0,1$)
is in agreement with the solution set of Eq.(\ref{Florides}). Anti-de
Sitter ($\Lambda<0$) and Milne ($R_{g}\neq const$) drop out as solutions
(since we have $E_{g}>0$ and $R_{g}=const$). Minkowski is a bit
more complicated, as it is a limiting case of de Sitter for $R_{g}\rightarrow\infty$,
so we defer a conclusion. Technically though, it is de Sitter, therefore
can be treated as such.

The intriguing part of this result is that we assumed the presence
of matter, $\rho_{m}>0$, nevertheless obtained ``empty'' de Sitter
as the unique solution. This suggests that de Sitter space is actually
non-empty. Moreover, from a relational perspective, inertia, energy,
space and time all emerge from the interaction of matter \citep{Mach,sep-RelationalTheories}.
So in this view empty spacetime cannot even exist (this was actually
subject of a long debate between Einstein and de Sitter, where Einstein
held this Machian view for a long time, but finally accepted emptiness
of de Sitter space \citep{CollectedPapersEinsteinVol8}). In section
\ref{Section:Grav time dilation} we infer that gravitational time
dilation makes de Sitter apparently empty. Misner-Sharp energy holds
for arbitrary spherically symmetric spacetimes, so the assumption
of the presence of matter is justified, even though this assumption
is rather implicit in Eq.(\ref{E MS}). Non-emptiness of de Sitter
also follows, quite explicitly, from the Machian derivation in section
\ref{Section:machian energy}. 

From Eqs.(\ref{RA-1},\ref{EAH}) we obtain, 

\begin{equation}
E_{g}=\frac{4}{3}\pi R_{g}^{3}(\rho-\rho_{k}),\label{E MS3}
\end{equation}
which is Eq.(\ref{E MS-1}) in the specific case of $R=R_{g}$ (in
section \ref{Section:Curvature of time} we find that curvature of
time renders curved de Sitter spatially flat, which explains the remarkable
flat volume of Misner-Sharp energy). Total energy $E_{g}$ may be
represented by the density $\rho_{E}$, 

\begin{equation}
E_{g}=\frac{4}{3}\pi R_{g}^{3}\rho_{E},\label{E MS2}
\end{equation}
hence, total density 
\begin{equation}
\rho=\rho_{E}+\rho_{k}.\label{rhotot-1}
\end{equation}
Given constancy of both $E_{g}>0$ and $R_{g}$, also
\begin{equation}
\rho_{E}=const>0,\label{rhoE}
\end{equation}
which therefore acts as constant vacuum energy density, i.e., 
\begin{equation}
\rho_{E}=\rho_{\Lambda},\label{rhoLambda}
\end{equation}
and so provides a natural interpretation of the cosmological constant.
Using Eqs. (\ref{RA-1},\ref{EAH},\ref{E MS2}) one obtains the following
expression of Misner-Sharp global energy, 
\begin{equation}
E_{g}=\frac{R_{g}^{3}}{2G}(H^{2}-\frac{8}{3}\pi G\rho_{k})=\frac{4}{3}\pi R_{g}^{3}\rho_{E}.\label{Eg}
\end{equation}
With Eq.(\ref{EAH}) this can be simplified to global energy per unit
mass $E\:(\equiv E_{g}/M_{g}=c^{2}=1)$, i.e., 
\begin{equation}
E=1=R_{g}^{2}(H^{2}-\frac{8}{3}\pi G\rho_{k})=\frac{8}{3}\pi GR_{g}^{2}\rho_{E}.\label{Eunit}
\end{equation}
This is the Friedmann equation multiplied on both sides by $R_{g}^{2}$,
which expresses the Friedmann equation as a global energy equation.
Note that the \textit{l.h.s.} of Eq.(\ref{Eunit}) adds an energy
constraint to the Friedmann equation, which, as we have seen, forces
$R_{g}$ and subsequently $\rho_{E}$ to be constant. For simplicity,
all energies hereafter are given per unit mass, so can be regarded
as potentials. 

Recalling that Misner-Sharp energy represents the sum of internal
kinetic energy and potential energy, one may recast equation (\ref{Eunit})
into the classical form $T=E-V$, where $T=R_{g}^{2}H^{2}$ is kinetic
energy and $V=-\frac{8}{3}\pi G\rho_{k}R_{g}^{2}=ka^{-2}$ is potential
energy due to curvature, i.e.,
\begin{equation}
T=H^{2}R_{g}^{2}=\frac{8}{3}\pi G(\rho_{E}+\rho_{k})R_{g}^{2}=1-ka^{-2}=E-V.\label{MS energy}
\end{equation}
This equation contains only global energies, $T$, $E$ and $V$.
Note that total density can be written
\begin{equation}
\rho=\rho_{E}+\rho_{k}=(1-ka^{-2})\rho_{E}.\label{tot density}
\end{equation}
Clearly, total density consists of only two components: total energy
$\rho_{E}=const$ and curvature energy $\rho_{k}=\rho_{E}a^{-2}$.
Where then has matter $\rho_{m}$ gone? According to Eq.(\ref{MS energy}),
$R_{g}^{2}H^{2}=1-ka^{-2}$, no matter what one assumes $\rho_{m}$
to be. However, turning this around: if the apparent horizon is to
meet conservation of total energy $E$, then this highly constrains
the equation of state of the density components to either $w=-1$
($\rho_{E}$), or $w=-\frac{1}{3}$ ($\rho_{k}$). Hence, since total
energy is represented by $\rho_{E}$, the conservation of Misner-Sharp
energy within the apparent horizon suggests that the \textit{energy
}density of total matter behaves as curvature energy, i.e.,
\begin{equation}
\rho_{m}=\rho_{k}.\label{eq:rhom=00003Drhok}
\end{equation}
This result seems at odds with the equation of state of individual
matter components, which agrees with the local notion of energy where
an isolated particle represents a certain energy. Eq.(\ref{eq:rhom=00003Drhok})
however regards the energy of total matter on the cosmological scale.
In section \ref{Section:non-locality of energy} we connect non-local
Misner-Sharp energy with a notion of relational energy between particles,
where an isolated point particle fundamentally does not represent
any energy at all. Instead, the interaction of the particle with its
cosmic background determines the energy associated with each particle.
Specifically, we find that the energy associated with matter is quite
different in an expanding universe than in peculiar motion in a fixed
background, where most of our physical notions and intuition come
from (see also section \ref{Section:Dark matter} on dark matter).

\subsection{Cosmic potential}

Both $E$ and $V$ are considered to contribute to the cosmic potential,
i.e., 
\begin{equation}
-\varphi\equiv E-V=\frac{8}{3}\pi G(\rho_{E}+\rho_{k})R_{g}^{2}=1-ka^{-2}.\label{phi-2}
\end{equation}
We see that in the curved case the potential evolves, which necessarily
induces gravitational time dilation. This feature is not recognized
in the standard FLRW metric, but has profound implications (see section
\ref{Section:Curvature of time}). Note that in the flat (free fall)
case the potential is constant, $\varphi=-c^{2}$, which agrees with
Sciama's cosmic potential \citep{Sciama}. The cosmic potential reminds
of the Newtonian potential at the center of the causal sphere,
\begin{equation}
\varphi_{N}=-2\pi G\rho R_{g}^{2},\label{PhiN}
\end{equation}
but the latter is off by a factor $\frac{3}{4}$, i.e., $\varphi_{N}=\frac{3}{4}\varphi$.
This points at an interesting relationship. Kinetic energy $T=H^{2}R_{g}^{2}$
is, for the appearance of $H$, naturally associated with recessional
motion of matter. However, Misner-Sharp energy $E$ also includes
energy of peculiar motion, which one may not immediately relate to
the Hubble parameter. That is unless the two, peculiar and recessional
energy, maintain a fixed ratio. This is what actually derives from
the Machian equation (section \ref{Section:machian energy}), which
expresses recessional and peculiar energy separately.

\subsection{Non-locality of energy\phantomsection\label{Section:non-locality of energy}}

In general relativity, notions of local and non-local energy coexist.
The non-local character of quasi-local Misner-Sharp energy fits the
relational view, where energy is exclusively a mutual property \textit{between}
causally connected particles. Any relational notion of energy is therefore
intimately connected with the density and size of the causal sphere,
i.e., with the presence of other matter. This view has become known
through Einstein as \textit{Mach's principle}, but actually dates
back to the relationalism of Descartes, Ockham or even Aristotle,
and has, apart from Mach and Einstein, been represented in particular
by Leibniz, Berkeley, Poincaré, Schr\"{o}dinger, Sciama, Brans and
Dicke \citep{sep-RelationalTheories,berkeley,BransDicke,Einstein1916,einstein1918prinzipielles,Mach,Schroedinger,Sciama,Barbour}.
Mach's principle (usually taken as: \textit{inertial properties arise
from the presence of cosmic mass}), fits into the relational view,
where inertia, energy, space and time emerge from the interaction
of particles (the terms \textit{relational} and \textit{Machian} have
a different historical context, but are largely exchangeable). In
the relational view, energy is a shared property, therefore not an
intrinsic property of a particle. This means that local energy in
fact does not exist in the relational universe; energy density is
merely global energy per unit volume. This may be understood realizing
that the potential energy of a particle of mass $m$ equals $m\varphi$,
where $\varphi$ is the cosmic potential. Without the cosmic mass-energy
present, the potential energy of the particle would vanish. A similar
argument applies to photon energy $h\nu$, where $\nu$ is the photon
frequency. A vanishing potential would redshift the photon frequency
to zero. Indeed, according to Eq.(\ref{EAH}), Misner-Sharp energy
within the apparent horizon equals the Schwarzschild mass. Hence the
idea that particle energy disappears in absence of other matter is
not uncommon. It can however be largely disregarded in a Universe
of constant potential, which is nothing but a flat empty Minkowski
background to local physics, where only local inhomogeneities of matter
are of interest. As Sciama noted, Newton's laws hold without referencing
any cosmic parameters \citep{Sciama}. 

Depending on $k$, the cosmic potential $-\varphi=1-ka^{-2}$ may
actually evolve, which changes the picture completely. Then the energy
associated with a particle appears to be explicitly related to the
presence of surrounding mass. In de Sitter, the comoving distance
to the apparent horizon evolves , $r_{g}(a)\propto a^{-1}$. The number
of particles contributing to the cosmic potential $\varphi$ depends
on the comoving distance to the horizon, therefore evolves too. In
light of this, it is conceivable that energy density not necessarily
dilutes as $a^{-3}$ in order for energy to be conserved. Moreover,
all different forms of mass-energy contribute to the cosmic potential,
and therefore to the energy associated with each particle, regardless
of species. Hence, the whole notion of separate energy densities of
distinct fluids is lost in the relational view. Instead, the cosmic
matter seems to behave as a single fluid with a uniform dilution rate.
This is a departure from standard cosmology, but not from general
relativity: uniform scaling is actually required to maintain (strict)
mass-energy equivalence, thus is in this sense a requirement within
general relativity. To the contrary, different scaling of radiation
and dust does violate both the mass-energy equivalence and the conservation
of energy. Indeed, as we have seen, uniform scaling follows from conservation
of Misner-Sharp energy within the apparent horizon.

Note that uniform scaling of matter density $\rho_{m}$ does not mean
that the actual \textit{number }densities of the different species
of particles maintain a fixed ratio. Particles may transform freely
from one form into another, so one still expects the number density
of photons to decay faster than the number density of massive particles,
due to the cool down of the Universe. For energy is conserved in such
conversions, these do not affect the total energy density of matter.
A single, uniformly scaling ``matter fluid'' does therefore not
conflict with the evolutionary picture of the Universe.

\subsection{Machian energy\phantomsection\label{Section:machian energy}}

Berkeley, an early critic of Newton, noted that one can not meaningfully
attribute a position or velocity to a single (point) particle in empty
space. Consequentially, this applies to kinetic energy too. He continues
noting that of two particles in otherwise empty space, only their
radial distance is observable \citep{berkeley}. Motion in any perpendicular
direction, like with these two particles in circular orbit of each
other, is unobservable in an empty background. Therefore (and this
is crucial), motion in non-radial direction, does \textit{not} represent
energy between two point particles. This means that both the kinetic
energy $T_{ij}$ and potential energy $V_{ij}$ between point particles
$i$ and $j$ depend \textit{only} on their separation $R_{ij}$ (or
time derivative thereof) \citep{Schroedinger,Barbour}. Actually,
Newtonian potential energy 
\begin{equation}
V_{ij}=-Gm_{i}m_{j}R_{ij}^{-1}\label{Vij}
\end{equation}
is perfectly Machian; it is indeed a mutual property between two connected
particles and depends geometrically only on their separation. Newtonian
kinetic energy, on the contrary, is defined relative to a frame of
reference, so is clearly not relational. Schr\"{o}dinger \citep{Schroedinger,Barbour}
reproduced Einstein's expression of the anomalous perihelion precession
from the following definition of Machian kinetic energy, 
\begin{equation}
T_{ij}={\textstyle \frac{1}{2}}\frac{V_{ij}}{\varphi_{p}}\dot{R}_{ij}^{2}.\label{Tij-1}
\end{equation}
The potential $\varphi_{p}$ is a scaling constant to match Newtonian
kinetic energy in peculiar motion \citep{TelkampPhysRevD.94.043520}.
Definition Eq.(\ref{Tij-1}) meets the Machian requirements: kinetic
energy $T_{ij}$ is mutual between two particles, is frame independent,
depends only on the radial component of motion, and vanishes if $V_{ij}\rightarrow0$.
As a consequence of the exclusively radial relationship, the mass
$m_{j}$ only contributes to the kinetic energy $T_{ij}$ if $\dot{R}_{ij}\neq0$.
This implies that only a part ($\varphi_{p}$) of the total Newtonian
potential $\varphi_{N}=-2\pi G\rho R_{g}^{2}$ adds to the Newtonian
inertia $m_{i}$ of a particle $i$ in peculiar motion. In a homogeneous,
isotropic sphere this part is $\bigl\langle\dot{R}_{ij}^{2}\bigr\rangle/\bigl\langle v_{ij}^{2}\bigr\rangle=\frac{1}{3}$,
where $v_{ij}$ is the relative speed, so that the effective potential
in peculiar motion is (cf. \citep{TelkampPhysRevD.94.043520}) 
\begin{equation}
\varphi_{p}={\textstyle \frac{1}{3}\varphi_{N}}.\label{Phip}
\end{equation}
Note that what we normally regard as inertial mass relates to peculiar
motion. Different from peculiar motion, recession is purely radial
motion between all particles, therefore the kinetic energy of recession
balances with the full potential, i.e., the potential in recessional
motion is 
\begin{equation}
\varphi_{r}=\varphi_{N}.\label{Phir}
\end{equation}
This however means that a particle in recessional motion effectively
has three times the inertial mass of the same particle in peculiar
motion. This is an intriguing consequence of Berkeley's conjectures,
which actually hints at the origin of unidentified ``dark'' matter,
as discussed in more detail in section \ref{Section:Dark matter}. 

Due to the frame independent formulation of $T_{ij}$, even a particle
at rest has kinetic energy associated with it. In this way, one can
express the recessional energies associated with a particle at rest
in the Hubble flow. Initially we shall ignore peculiar motion. Adopting
the elementary definition Eq.(\ref{Tij-1}), integration over the
cosmic comoving volume $\mathcal{V}_{g}=\mathcal{V}(r_{g})$ yields
the recessional Machian kinetic energy $T_{r}$, between a unit mass
test particle at rest in the Hubble flow and all receding masses within
the horizon \citep{TelkampPhysRevD.94.043520},

\begin{equation}
T_{r}=\underset{\mathcal{V}_{g}}{\intop}\:\tfrac{1}{2}\!\frac{\textrm{d}\varphi_{r}(r,\theta,\phi)}{\tfrac{1}{3}\varphi_{r}}\,r^{2}\dot{a}^{2}={\textstyle \frac{3}{4}}r_{g}^{2}\dot{a}^{2}={\textstyle \frac{3}{4}}H^{2}R_{g}^{2},\label{eq:T-1}
\end{equation}
where $\textrm{d}\varphi_{r}(r,\theta,\phi)$ represents the potential
at the origin due to the receding infinitesimal cosmic volume element
at spherical coordinates ($r,\theta,\phi$). According to Eqs.(\ref{PhiN},\ref{Phir}),
the potential in recessional motion is the Newtonian potential
\begin{equation}
\varphi_{r}=-2\pi G(\rho_{E}+\rho_{m})R_{g}^{2},\label{phi-1}
\end{equation}
where we explicitly assume the presence of matter $\rho_{m}$ next
to total energy $\rho_{E}$. Eqs.(\ref{MS energy},\ref{eq:T-1})
show that recessional kinetic energy is $\frac{3}{4}$ of total kinetic
energy, thus accordingly $\varphi_{r}=\frac{3}{4}\varphi=-\frac{3}{4}(1-ka^{-2})$.
With Eqs.(\ref{eq:T-1},\ref{phi-1}) this gives the Machian equation
of recessional energy per unit mass 
\begin{equation}
T_{r}={\textstyle \frac{3}{4}}H^{2}R_{g}^{2}=2\pi G(\rho_{E}+\rho_{m})R_{g}^{2}={\textstyle \frac{3}{4}}(1-ka^{-2})=E_{r}-V_{r}.\label{Mach rec}
\end{equation}
Recalling that the effective potential in peculiar motion is only
$\frac{1}{3}$ of the potential in recessional motion, $\varphi_{p}=\frac{1}{3}\varphi_{r}$,
we expect the balancing kinetic energies to maintain the same ratio,
i.e., the kinetic energy of peculiar motion per unit mass is $T_{p}=\frac{1}{3}T_{r}$,
therefore
\begin{equation}
T_{p}={\textstyle \frac{1}{4}}H^{2}R_{g}^{2}={\textstyle \frac{2}{3}}\pi G(\rho_{E}+\rho_{m})R_{g}^{2}={\textstyle \frac{1}{4}}(1-ka^{-2})=E_{p}-V_{p}.\label{Mach pec}
\end{equation}
The Machian recessional and peculiar energy combined add to 
\begin{equation}
T=H^{2}R_{g}^{2}={\textstyle \frac{8}{3}}\pi G(\rho_{E}+\rho_{m})R_{g}^{2}=1-ka^{-2}=E-V.\label{Mach energy}
\end{equation}

\subsection{Cross-interpretation\phantomsection\label{Section:cross interpretation}}

The Machian equation Eq.(\ref{Mach energy}) is the same as Eq.(\ref{MS energy})
for Misner-Sharp energy, except that $\rho_{m}$ in the Machian case
takes the place of $\rho_{k}$ in the Misner-Sharp case. In both equations
total energy and kinetic energy have the same meaning and the same
value, i.e., $E=T+V=1$, and $T=H^{2}R_{g}^{2}$. So the potential
energy $V$ in both equations must be identical too. Stated otherwise,
total density in both expressions is equal,
\begin{equation}
\rho=\rho_{E}+\rho_{k}=\rho_{E}+\rho_{m},\label{rhotot}
\end{equation}
hence $\rho_{k}=\rho_{m}$, in agreement with the earlier conclusion
of Eq.(\ref{eq:rhom=00003Drhok}). That is, subject to conservation
of global energy, matter behaves as curvature energy, which identifies
matter as source of curvature energy. For we demand the presence of
matter, i.e., $\rho_{m}>0$, this seems to single out open de Sitter
($\rho_{k}>0$) as the unique solution. Since $\rho=\rho_{E}+\rho_{m}>0$,
Minkowski is excluded as a possible limiting case of open de Sitter.

The particularity of curvature energy is that it represents extrinsic
curvature; it appears in accelerated frames and vanishes in the free
fall frame, revealing flat de Sitter. Even so, we do not expect matter
energy $\rho_{m}$ (nor $\rho_{k}$) to actually nullify, for it represents
energy. Indeed, nullification of $\rho_{m}$ implies $\rho_{E}=\rho_{m0}=0$,
which contravenes Eq.(\ref{rhoE}). As will follow, the vanishing
of matter/curvature in free fall is only apparent, as an effect of
gravitational time dilation. 

\subsection{Gravitational time dilation\phantomsection\label{Section:Grav time dilation}}

Having established $\rho_{k}=\rho_{m}>0$, we assume hereafter that
the Universe is negatively curved and can be represented by the open
de Sitter spacetime. According to Eq.(\ref{phi-2}), curvature $k=-1$
implies evolution of the potential, which induces gravitational time
dilation, i.e.,
\begin{equation}
dt^{2}=g_{t't'}dt'^{2},\label{time dilation}
\end{equation}
where $t$ is the unaccelerated cosmic time. The prime denotes the
accelerated time coordinate $t'$ in open de Sitter, which we shall
identify below as \textit{observational time coordinate} on the past
light cone. Unaccelerated cosmic time is associated with flat de Sitter,
where $\rho=\rho_{E}$ and $\varphi=1$. Hence, in the freely falling
comoving frame, the curved open de Sitter universe is transformed
into flat de Sitter. Note that free fall in this context specifically
concerns the time dimension; the observer in ($r$,$t'$) coordinates
is accelerated in the time dimension and is therefore not in free
fall, even though spatially at rest in the Hubble flow like the unaccelerated
freely falling comoving observer in ($r$,$t$) coordinates. This
is analog to \citep{PARIKH2002189}, where Parikh shows the transformation
of coordinate time in static de Sitter to the Painlevé-de Sitter time
coordinate of the free fall observer. Here, the mechanism of transformation
from open de Sitter to flat de Sitter is through transformation of
total density by gravitational time dilation, as follows. 

In terms of the curved time coordinate $t'$, the Friedmann equation
of open de Sitter ($k=-1$) is

\begin{equation}
\frac{1}{a^{2}}\frac{da^{2}}{dt'^{2}}\propto\rho_{E}+\rho_{k}=(1-ka^{-2})\rho_{E}.\label{da/dt'}
\end{equation}
Referencing definition Eq.(\ref{time dilation}), in free fall Eq.(\ref{da/dt'})
transforms to flat de Sitter ($k=0$),
\begin{equation}
\frac{1}{a^{2}}\frac{da^{2}}{dt^{2}}\propto\frac{\rho_{E}+\rho_{k}}{g_{t't'}}=\rho_{E},\label{da/dt}
\end{equation}
so that indeed curvature/matter energy seems to vanish due to the
effect of time dilation, 

\begin{equation}
g_{t't'}=\frac{\rho_{E}+\rho_{k}}{\rho_{E}}=1-ka^{-2}.\label{gt't'-1}
\end{equation}
Hence, total density in free fall equals $\rho_{E}$, as expected,
but still consists underlying of both transformed components, as is
evident from Eq.(\ref{da/dt}). It is this property of spacetime which
turns the evolving potential $\varphi=1-ka^{-2}$ in free fall to
the constant $\varphi=1$, thereby hides matter/curvature energy.
Thus also time dilation explains why de Sitter is only apparently
empty. Furthermore, the constant density $\rho=\rho_{E}$ falsely
suggests non-evolution. Inspection of Eq.(\ref{da/dt}) shows that
the contribution of matter/curvature to constant density equals 
\begin{equation}
\frac{\rho_{k}}{g_{t't'}}=\frac{\rho_{E}a^{-2}}{1+a^{-2}}=\frac{\rho_{E}}{a^{2}+1},\label{rhok'}
\end{equation}
which scales down to zero as $a\rightarrow\infty$. The contribution
of total energy, on the other hand, ramps up asymptotically to $\rho_{E}$
according to 
\begin{equation}
\frac{\rho_{E}}{g_{t't'}}=\frac{\rho_{E}}{1+a^{-2}}.\label{rhoE'}
\end{equation}

Flat de Sitter of course reminds of the non-evolutionary \textit{steady
state theory} \citep{BondiGold,Hoyle} by Bondi, Gold and Hoyle, which
however features continuous creation of matter to maintain constant
density, instead of gravitational time dilation in the present cosmology. 

The identification of matter with curvature energy is reminiscent
of \textit{K-matter} \citep{kolb1989coasting} and Milne cosmology.
It fits the idea of coasting galaxies in the homogeneous, zero-field
universe, as pointed out, e.g., by Layzer \citep{layzer}. Here, however,
matter/curvature density is always in the company of constant total
energy density, i.e., $\rho_{E}=a^{2}\rho_{m}$, yielding exclusively
open/flat de Sitter.

\subsection{Cosmological solutions\phantomsection\label{Section:cosmol solutions}}

From the foregoing we conclude that conservation of global energy
in a non-empty universe is consistent with open de Sitter in terms
of the accelerated time coordinate $t'$, 
\begin{equation}
\frac{1}{a^{2}}\frac{da^{2}}{dt'^{2}}=H_{\varLambda}^{2}\left(1+a^{-2}\right),\label{open dS-1}
\end{equation}
where $H_{\varLambda}^{2}\equiv c^{2}/R_{g}^{2}$. This solution transforms
in free fall to flat de Sitter, in terms of unaccelerated cosmic time
$t$,
\begin{equation}
\frac{1}{a^{2}}\frac{da^{2}}{dt^{2}}=H_{\varLambda}^{2}.\label{flat dS}
\end{equation}

\section{Observational validity?\phantomsection\label{Section:Obs.val.}}

Open/flat de Sitter is the unique solution of conservation of global
energy. Nevertheless, observation seems to rule out both representations.
CMB data indicate that our Universe is spatially flat, which seems
to leave flat de Sitter as the single option. Supernova observations,
however, clearly disfavor flat de Sitter relative to \textgreek{L}CDM.
On the other hand, the open de Sitter model Eq.(\ref{open dS-1}),
is close to \textgreek{L}CDM in expansion dynamics. It has a deceleration
$q_{0}=-\frac{1}{2}$ at present epoch, quite close to \textgreek{L}CDM
with $q_{0}\sim-0,55$ ($\Omega_{m}\sim0.3$, $\Omega_{\Lambda}\sim0.7$).
But obviously, open de Sitter, as derived from the standard metric,
is spatially curved, therefore is not supported by CMB data. The two
de Sitter models thus seem to match different sets of observations
quite well, but neither fits both. The situation however changes once
gravitational time dilation is taken into account. 

\subsection{Curvature of time vs. curvature of space\phantomsection\label{Section:Curvature of time}}

Gravitational time dilation is not accommodated for in the standard
FLRW metric, 
\begin{equation}
ds^{2}=dt{}^{2}-a^{2}\frac{dr^{2}}{1-kr^{2}}-a^{2}r^{2}d\varOmega^{2},\label{FLRW-3}
\end{equation}
where $g_{tt}=1=const$. Evolution of the potential thus prompts adjustment
of the metric as follows. 

The metric to the free fall flat de Sitter model is the flat standard
metric, 
\begin{equation}
ds^{2}=dt^{2}-a^{2}dr^{2}-a^{2}r^{2}d\varOmega^{2}.\label{Flat metric}
\end{equation}
As pointed out in section \ref{Section:Grav time dilation}, open
de Sitter and its free fall representation, flat de Sitter, only differ
by their time coordinates due to gravitational time dilation, according
to Eq.(\ref{gt't'-1}). The metric Eq.(\ref{Flat metric}) in terms
of the accelerated time coordinate $t'$ is 
\begin{equation}
ds^{2}=(1-ka^{-2})dt'{}^{2}-a^{2}dr^{2}-a^{2}r^{2}d\varOmega^{2},\label{cosmic metric-1}
\end{equation}
where time is curved, but space is flat. Thus index $k$ represents
the extrinsic curvature of time instead of space. An insightful way
to retrieve open de Sitter is to evaluate the metric Eq.(\ref{cosmic metric-1})
on the light cone ($ds=0$) at the horizon ($r=r_{g}\propto a^{-1}$)
and setting $k=-1$, while using $r_{g}da+adr_{g}=0$. This returns
the Friedmann equation of open de Sitter, though derived from a spatially
flat metric, Eq.(\ref{cosmic metric-1}). This means that curvature
of the time dimension accounts completely for the extrinsic curvature
of open de Sitter. As a result, open de Sitter is necessarily spatially
flat, which makes this model observationally viable. Furthermore,
the derivation of open de Sitter from the metric on the light cone
identifies the accelerated coordinate $t'$ as observational time
coordinate on the past light cone. 

Note that gravitational time dilation gives an explanation to why
the Misner-Sharp mass in Eq.(\ref{E MS2}) coincides with a flat volume,
regardless of curvature $k$. 

\subsection{Relational physics of missing matter\phantomsection\label{Section:Dark matter}}

The relative density of baryonic matter is currently estimated at
$\widetilde{\Omega}_{b}\sim0.05$ \citep{Planck2015}. In the \textgreek{L}CDM
model the relative density of dust is $\widetilde{\Omega}_{d}\sim0.30$,
so that the deficit of matter is $\widetilde{\Omega}_{c}\sim0.25$,
which is attributed to dark matter. The relative density of matter
in the open de Sitter model of Eq.(\ref{open dS-1}) is $\Omega_{m}=\frac{1}{2}$,
meaning that an even larger fraction $\Omega_{c}\sim0.45$ of matter
energy is to be identified. Recall that density $\Omega_{m}$ regards
the global potential energy associated with matter, not the local
energy density of particles. As pointed out in section \ref{Section:machian energy},
Berkeley's ontological notions on radial motion entail a three times
larger inertial mass of a particle in (the purely radial) recessional
motion as compared with the same particle in peculiar motion. Our
unit of inertial mass relates to peculiar motion, therefore inertial
mass gets boosted in recessional motion. An additional effect, due
to time dilation, is that at present time the densities in the accelerated
time coordinate $t'$ are twice the densities in the free fall cosmic
time coordinate. This doubles the present energy density of matter
in the observational coordinates of the open de Sitter model. The
ratio of inertial mass in the different cases is reflected in the
ratio of the corresponding kinetic energies as follows. According
to Eq.(\ref{Mach pec}), in peculiar motion the kinetic energy per
unit mass equals $T_{p}={\textstyle \frac{1}{4}}(1-ka^{-2})$. Thus
in the free fall frame ($k=0$) of flat de Sitter this is $T_{p}={\textstyle \frac{1}{4}}$.
This is the frame in which we define our standard unit of inertial
mass in peculiar motion. The kinetic energy in recessional motion
equals according to Eq.(\ref{Mach rec}) $T_{r}={\textstyle \frac{3}{4}}(1-ka^{-2})$,
therefore $T_{r}={\textstyle \frac{3}{4}}$ in free fall. The combined
peculiar and recessional motion in the same frame amounts to $T=T_{p}+T_{r}={\textstyle \frac{1}{4}}+\frac{3}{4}=1$.
Finally, transformation from cosmic time to the accelerated observational
time coordinate $t'$ in open de Sitter doubles the kinetic energy
at present time to $T'_{0}=2$. Thus the standard unit of mass in
peculiar motion in flat de Sitter gets multiplied by a factor $T'_{0}/T_{p}=8$
in the observational open de Sitter model. The present cosmology therefore
predicts a relative density of baryonic matter (in standard units
related to peculiar motion) of exactly $\Omega_{b}=\Omega_{m}/8=1/16=0.0625$,
which is quite close to current estimates of $\sim0.05$. Though,
the global energy density associated with this baryonic matter in
open de Sitter is $\frac{1}{2}.$ This is relational physics.

\subsection{Model comparison using SNIa data\phantomsection\label{Section:Model comp}}

The physics involved in the present cosmology (in particular gravitational
time dilation instead of spatial curvature) must be accounted for
accordingly in the physical models underlying cosmological probes
such as CMB, BAO and SNIa. Hence, comparison with models of standard
cosmology is not necessarily on the same premises. Open de Sitter
in standard cosmology implies spatial curvature, whereas time-dilated
open de Sitter (open dS') is spatially flat. This difference impacts
cosmological calculations, e.g., of the transverse comoving distance,
even while the two models have the exact same Friedmann equation.
In the case of SNIa, this difference can be accounted for straightforwardly.
The physics of CMB and BAO is more involved and requires deeper analysis,
beyond present scope. Thus the following model comparison is only
indicative, since it is limited to SNIa.

\smallskip{}

The Union2.1 compilation of $n=580$ SNIa samples from the \textit{Supernova
Cosmology Project} (\textit{SCP}) \citep{SupernovaCosmologyProject2012ApJ}
gives observed distance modulus $\widetilde{\mu}(z)$ versus redshift
$z$. We use this data set without any processing. Following Suzuki
\textit{et al}. \citep{SupernovaCosmologyProject2012ApJ}, we minimize
\begin{equation}
\chi^{2}=(\widetilde{\mu}-\mu)'C^{-1}(\widetilde{\mu}-\mu),\label{X2}
\end{equation}
where $\mu(z)$ is the model prediction of $\widetilde{\mu}$, and
$C$ is a diagonal weighing matrix of sample error variances without
systematics, as provided by \textit{SCP}. The prediction is the theoretical
value of the distance modulus, given (in $Mpc$) by $\mu(z)=m(z)-M=5log_{10}(D_{L})+25$.
The luminosity distance $D_{L}$ relates according to $D_{L}=(z+1)D_{M}$
to the transverse comoving distance $D_{M}$, while $D_{M}$ in standard
cosmology relates to the comoving distance $D_{c}$, depending on
spatial curvature index $k$, according to 
\begin{equation}
D_{M}=\frac{D_{H}}{\sqrt{\Omega_{k}}}S\Bigl(\frac{\sqrt{\Omega_{k}}}{D_{H}}D_{c}\Bigr),\label{Dtransv}
\end{equation}
where
\begin{equation}
S(x)=\begin{cases}
sinh(x) & k=-1\\
x & k=0\\
sin(x) & k=1.
\end{cases}\label{S(x)}
\end{equation}
In the present cosmology, extrinsic curvature only regards the time
dimension; space is manifestly flat. That is, $D_{M}=D_{c}$, regardless
of $k$. Finally, the comoving distance $D_{c}$ follows from
\begin{equation}
D_{c}(z)=\underset{0}{\intop^{{\scriptstyle {\scriptscriptstyle Z}}}}\frac{dz'}{H(z')}.\label{Dc}
\end{equation}

The models considered are: flat \textgreek{L}CDM, open dS', power
law, Milne and flat de Sitter. The results of minimizing $\chi^{2}$
are summarized in Tables \ref{tab:Model best fit} and \ref{tab:Model-performance},
which show model structure, the number of free parameters $n_{p}$,
best fit parameter values, and the values of different model performance
criteria (the best score per column in boldface). The models are sorted
by $\chi^{2}$ score. As usual, \textgreek{L}CDM performs best on
$\chi^{2}$. Open dS', however, fits the data nearly as well, while
it has only one degree of freedom $n_{p}=1$, against $n_{p}=2$ for
\textgreek{L}CDM. This is rewarded by the significantly better BIC
scores of open dS', while \textgreek{L}CDM remains preferable according
to AIC scores. The $\textrm{w}_{i}(\textrm{BIC)}$ and $\textrm{w}_{i}(\textrm{AIC)}$
are estimated model probabilities, which allow for somewhat easier
interpretation than $\varDelta$BIC and $\varDelta$AIC scores \citep{Wagenmakers2004}.
From BIC and AIC together one may slightly favor open dS', but given
the noise levels of the data, the difference is too small to be conclusive.
Moreover, the purpose here is not to decide on best model, but to
point at open dS' as a viable alternative of \textgreek{L}CDM. 

The power law model does rather well, but is purely empirical; it
lacks physical motivation. Milne and flat de Sitter are cosmologically
relevant (being part of Florides' model set), but perform relatively
poorly. These models however support the conclusion that open dS'
and \textgreek{L}CDM are quite close. 

\begin{table}
\caption{\label{tab:Model best fit}Model parameter estimates}

\begin{tabular}{cccc||c||c||cr@{\extracolsep{0pt}.}lr@{\extracolsep{0pt}.}lr@{\extracolsep{0pt}.}l}
\hline 
 & $H^{2}/H_{0}^{2}$ &  & \multicolumn{4}{c}{$\Omega,\,q$} & \multicolumn{2}{c}{} & \multicolumn{2}{c}{$H_{0}$} & \multicolumn{2}{c}{}\tabularnewline
\hline 
\hline 
 \textgreek{L}CDM & $\Omega_{d}a^{-3}+\Omega_{\Lambda}$ &  & \multicolumn{4}{c}{$\Omega_{d}=$0.278,$\:\Omega_{\Lambda}=1-\Omega_{d}$ } & \multicolumn{2}{c}{} & 70&0 & \multicolumn{2}{c}{}\tabularnewline
open dS' & $\Omega_{m}a^{-2}+\Omega_{\Lambda}$ &  & \multicolumn{4}{c}{$\Omega_{m}=\Omega_{\Lambda}=\frac{1}{2}$ } & \multicolumn{2}{c}{} & 69&9 & \multicolumn{2}{c}{}\tabularnewline
power law & $a^{-2-2q}$ &  & \multicolumn{4}{c}{$q=-0.36$ $\rightarrow a(t')=t'^{1.56}$} & \multicolumn{2}{c}{} & 69&3 & \multicolumn{2}{c}{}\tabularnewline
Milne & $\Omega_{k}a^{-2}$ &  & \multicolumn{4}{c}{$\Omega_{k}=1$ } & \multicolumn{2}{c}{} & 66&6 & \multicolumn{2}{c}{}\tabularnewline
flat dS & $\Omega_{\Lambda}$ &  & \multicolumn{4}{c}{$\Omega_{\Lambda}=1$} & \multicolumn{2}{c}{} &  74&5 & \multicolumn{2}{c}{}\tabularnewline
\hline 
\end{tabular}
\end{table}
\begin{table}
\caption{Model performance\label{tab:Model-performance}}
\begin{tabular}{ccr@{\extracolsep{0pt}.}lr@{\extracolsep{0pt}.}lr@{\extracolsep{0pt}.}lr@{\extracolsep{0pt}.}lr@{\extracolsep{0pt}.}lr@{\extracolsep{0pt}.}lr@{\extracolsep{0pt}.}lr@{\extracolsep{0pt}.}l}
\hline 
 &  & \multicolumn{2}{c}{$\chi^{2}/n$} & \multicolumn{2}{c}{$n_{p}$} & \multicolumn{2}{c}{} & \multicolumn{2}{c}{$\varDelta$BIC} & \multicolumn{2}{c}{w$_{i}$(BIC)} & \multicolumn{2}{c}{} & \multicolumn{2}{c}{$\varDelta$AIC} & \multicolumn{2}{c}{w$_{i}$(AIC)}\tabularnewline
\hline 
\hline 
\textgreek{L}CDM &  & \textbf{0}&\textbf{9694 } & \multicolumn{2}{c}{2 } & \multicolumn{2}{c}{} & 3&22  & 0&16  & \multicolumn{2}{c}{} & \multicolumn{2}{c}{\textbf{0}\textit{ }} & \textbf{0}&\textbf{62}\textit{ }\tabularnewline
open dS' &  & 0&9746  & \multicolumn{2}{c}{1 } & \multicolumn{2}{c}{} & \multicolumn{2}{c}{\textbf{0 }} & \textbf{0}&\textbf{82 } & \multicolumn{2}{c}{} & 1&14  & 0&35\tabularnewline
power law &  & 0&9798 & \multicolumn{2}{c}{2 } & \multicolumn{2}{c}{} & 9&46  & 7&3E-3  & \multicolumn{2}{c}{} & 6&24 & 0&027\tabularnewline
Milne &  & 1&0404  & \multicolumn{2}{c}{1 } & \multicolumn{2}{c}{} & 37&9  & 4&9E-9  & \multicolumn{2}{c}{} & 39&0 & 2&1E-9\tabularnewline
flat dS &  & 1&6481  & \multicolumn{2}{c}{1 } & \multicolumn{2}{c}{} & 304&7  & 5&7E-67 & \multicolumn{2}{c}{} & 305&9 & 2&4E-67 \tabularnewline
\hline 
\end{tabular}
\end{table}

\section{Theoretical evaluation\phantomsection\label{Section:Theor eval}}

Theoretical issues of concordance cosmology, or standard cosmology
in general, are revisited for the present cosmology as follows.
\begin{enumerate}
\item \textit{Horizon problem}. The smoothness of the CMB is difficult to
explain without divergence of the particle horizon, by which all matter
is causally connected at initial time. \textgreek{L}CDM lacks this
property due to deceleration $q>0$ at early times. A short flat de
Sitter inflation phase at initial time is hypothesized to provide
the causal connection of all matter at initial time. The present de
Sitter solution has accelerated expansion from initial time onward,
i.e., $q\leq0$, therefore has a diverging particle horizon at initial
time, which provides smoothness of the CMB, without inflation. 
\item \textit{Flatness problem}. Flat space is an unstable equilibrium of
the \textgreek{L}CDM model; any small deviation from a perfectly flat
\textgreek{L}CDM universe causes density to run away from critical
density. Inflation creates an extremely flat initial state of the
\textgreek{L}CDM universe, as a way to delay notable curvature. The
flatness problem does not apply to the present cosmology; the spatial
part of the metric Eq.(\ref{cosmic metric-1}) is manifestly flat,
even if $k\neq0$, since all extrinsic curvature is (necessarily)
in the time dimension. 
\item \textit{Coincidence problem}. The densities $\rho_{E}$ and $\rho_{m}$
$(=\rho_{E}a^{-2})$ in open de Sitter are obtained analytically and
are exactly equal at present time $t'_{0}$. Thus coincidence of nearly
equal densities at present time, as in \textgreek{L}CDM, is absent.
Flat de Sitter obviously has no coincidence problem either.
\item \textit{Unidentified dark matter}. According to section \ref{Section:Dark matter},
a density of baryonic matter $\Omega_{b}$ in the standard definition
accounts for an effective \textit{energy density} of baryonic matter
$\Omega_{m}=8\Omega_{b}$ in open de Sitter. Since $\Omega_{m}=\frac{1}{2}$
in open de Sitter, this predicts a baryonic density of $\Omega_{b}=0.0625$,
which is close to present estimates of around $\widetilde{\Omega}_{b}=0.05$
\citep{Planck2015}. This result follows specifically from relational
considerations. 
\item \textit{Unidentified dark energy}. The density of Misner-Sharp total
internal energy within the de Sitter horizon is identified as a cosmological
constant, i.e., $\rho_{E}=\rho_{\Lambda}=const$. The global Misner-Sharp
energy per unit mass amounts to $E=1$ and equals the sum of total
energy of recessional ($E_{r}=\frac{3}{4}$) and peculiar motion ($E_{p}=\frac{1}{4}$),
as derived from the Machian energy equations, Eqs.(\ref{Mach rec},\ref{Mach pec}). 
\item \textit{Age problem}. Since $a(t')=\sinh(H_{\Lambda}t')$, the age
of the open de Sitter Universe is $t'_{0}=H_{\Lambda}^{-1}\textrm{asinh}(1)=\sqrt{2}H_{0}^{-1}\textrm{asinh}(1)\sim1.25H_{0}^{-1}\sim17.9$
billion years (at $H_{0}=68$ km/s/Mpc), which is a comfortable high
age, given constraints near $H_{0}^{-1}\sim14.4$ billion years. Estimated
ages of the oldest known objects are around the predicted age of 13.8
billion years of the \textgreek{L}CDM universe \citep{Planck2015}. 
\item \textit{Violation of the conservation of energy}. The present cosmology
is derived from conservation of quasi-local Misner-Sharp energy within
the causal sphere, thus necessarily satisfies conservation of this
global energy. The result agrees with the Machian expression of total
energy. 
\item \textit{Violation of the (strict) mass-energy equivalence principle.}
In standard cosmology this violation is due to different scaling of
the energy densities of different forms of matter. In Machian context,
energy is a mutual property between connected particles of mixed sorts,
the total effect of which is aggregated into the cosmic potential.
Hence all forms of matter contribute to the energy associated with
each particle, thus total matter behaves as a single fluid. The present
analysis shows that, subject to conservation of Misner-Sharp global
energy, the total energy density of (an arbitrary mixture of) matter
behaves as curvature energy density. How uniform scaling of potential
energy of total matter relates to the different equation of state
of particles in local physics remains to be answered. The present
cosmology suggests that the different equation of state of different
forms of matter rule the evolution of number densities of these forms,
but not the potential energy of total matter. 
\item \textit{Unclear relationship of sources and curvature energy}. The
present cosmology points at identity of the matter energy and curvature
energy, i.e., $\rho_{m}=\rho_{k}$.
\item \textit{Coexistence of different ``causal'' horizons at different
distances}. In the de Sitter universe the event horizon and apparent
horizon coincide, so the ambiguous causal status of particles in between
these horizons disappears.
\item \textit{Incompatible forms of redshift}. There are compelling reasons
to assume that cosmological, gravitational and Doppler redshift are
all three relevant to cosmology. Even so, a long history of controversy
about the subject has not resolved the apparent inconsistency of these
different forms of redshift in Friedmann universes \citep{Chodorowski:2009cb,Faraoni:2009buREDSHIFT,Harrison}.
Without going into this debate, one may note that the three only need
to be consistent in the unique spacetime of our Universe. Melia \citep{Melia:2012ic}
showed that the three forms of redshift are consistent in all six
constant curvature spacetimes of Eq.(\ref{Florides}), which includes
de Sitter. 
\item \textit{Representation of physical time?} Conservation of Misner-Sharp
energy yields two consistent cosmological pictures of the Universe:
the evolutionary open dS' universe, as observable on the past light
cone, and the (apparent) steady state flat de Sitter universe of the
unaccelerated free fall observer, each representation with its own
time parameter, which is the only difference between the two. Which
time is our clock time? In standard cosmology, cosmic time in flat
de Sitter represents comoving clock time, which presumes constancy
of the potential. This indeed holds in free fall coordinates. The
comoving observer would decide that in flat de Sitter nothing changes
and that the Universe must be infinitely old. On the other hand, the
presence of matter in expanding flat de Sitter points at evolution,
which is confirmed by observation. In the present cosmology, evolution
goes together with a declining potential $-\varphi=1+a^{-2}$, causing
acceleration of clock rate relative to cosmic time, which suggests
that a comoving clock indicates accelerated time $t'$, rather than
cosmic time $t$. 
\end{enumerate}

\section{Conclusion}

Conservation of energy underlies many, if not all, physical laws and
relationships. One therefore does not expect a cosmology in violation
of energy conservation to be consistent, unambiguous and explicable.
We therefore considered the theoretical aspects of a cosmology derived
from conservation of Misner-Sharp global energy, leading to a single
cosmological solution without the conceptual problems of concordance
cosmology, as summarized in the previous section. Observational validity
of the model is only indicative, since limited to SNIa, where open
dS' and \textgreek{L}CDM perform equally well. 

Standard cosmology assumes a local notion of energy, admits different
equation of state of matter, and considers de Sitter to be empty.
In contrast, the relational view of physics assumes an exclusively
global notion of energy, considers a single total matter fluid of
uniform equation of state, and precludes the existence of empty space.
While this seems at variance with general relativity, conservation
of Misner-Sharp energy within the apparent horizon of a non-empty
universe yields de Sitter as the unique solution, with energy density
of total matter $\rho_{m}=\rho_{k}$, while total energy density $\rho_{E}$
acts as cosmological constant. The same results follow from a Machian
model. This shows, quite in the original spirit of Einstein \citep{Einstein1916},
that global energy in general relativity is consistent with relational
notions of energy, in particular Berkeley's ontology of the exclusively
radial relationship between particles. The relational definitions
allow for a more explicit treatment, leading to additional results,
like the quantification of peculiar and recessional energy, and a
predicted baryon density of exactly $\Omega_{b}=\frac{1}{8}\Omega_{m}=0.0625$,
which accounts for an effective global energy density of total matter
of $\Omega_{m}=\frac{1}{2}$ in open dS'. This emphasizes the relational
aspect of energy. 

\bibliographystyle{apsrev4-1}

\begin{thebibliography}{32}%
\makeatletter
\providecommand \@ifxundefined [1]{%
 \@ifx{#1\undefined}
}%
\providecommand \@ifnum [1]{%
 \ifnum #1\expandafter \@firstoftwo
 \else \expandafter \@secondoftwo
 \fi
}%
\providecommand \@ifx [1]{%
 \ifx #1\expandafter \@firstoftwo
 \else \expandafter \@secondoftwo
 \fi
}%
\providecommand \natexlab [1]{#1}%
\providecommand \enquote  [1]{``#1''}%
\providecommand \bibnamefont  [1]{#1}%
\providecommand \bibfnamefont [1]{#1}%
\providecommand \citenamefont [1]{#1}%
\providecommand \href@noop [0]{\@secondoftwo}%
\providecommand \href [0]{\begingroup \@sanitize@url \@href}%
\providecommand \@href[1]{\@@startlink{#1}\@@href}%
\providecommand \@@href[1]{\endgroup#1\@@endlink}%
\providecommand \@sanitize@url [0]{\catcode `\\12\catcode `\$12\catcode
  `\&12\catcode `\#12\catcode `\^12\catcode `\_12\catcode `\%12\relax}%
\providecommand \@@startlink[1]{}%
\providecommand \@@endlink[0]{}%
\providecommand \url  [0]{\begingroup\@sanitize@url \@url }%
\providecommand \@url [1]{\endgroup\@href {#1}{\urlprefix }}%
\providecommand \urlprefix  [0]{URL }%
\providecommand \Eprint [0]{\href }%
\providecommand \doibase [0]{http://dx.doi.org/}%
\providecommand \selectlanguage [0]{\@gobble}%
\providecommand \bibinfo  [0]{\@secondoftwo}%
\providecommand \bibfield  [0]{\@secondoftwo}%
\providecommand \translation [1]{[#1]}%
\providecommand \BibitemOpen [0]{}%
\providecommand \bibitemStop [0]{}%
\providecommand \bibitemNoStop [0]{.\EOS\space}%
\providecommand \EOS [0]{\spacefactor3000\relax}%
\providecommand \BibitemShut  [1]{\csname bibitem#1\endcsname}%
\let\auto@bib@innerbib\@empty
\bibitem [{\citenamefont {Florides}(1980)}]{Florides}%
  \BibitemOpen
  \bibfield  {author} {\bibinfo {author} {\bibfnamefont {P.}~\bibnamefont
  {Florides}},\ }\href@noop {} {\bibfield  {journal} {\bibinfo  {journal} {Gen
  Relat Gravit}\ }\textbf {\bibinfo {volume} {12}},\ \bibinfo {pages} {563}
  (\bibinfo {year} {1980})}\BibitemShut {NoStop}%
\bibitem [{\citenamefont {Misner}\ and\ \citenamefont
  {Sharp}(1964)}]{MisnerSharp}%
  \BibitemOpen
  \bibfield  {author} {\bibinfo {author} {\bibfnamefont {C.~W.}\ \bibnamefont
  {Misner}}\ and\ \bibinfo {author} {\bibfnamefont {D.~H.}\ \bibnamefont
  {Sharp}},\ }\href {\doibase 10.1103/PhysRev.136.B571} {\bibfield  {journal}
  {\bibinfo  {journal} {Phys. Rev.}\ }\textbf {\bibinfo {volume} {136}},\
  \bibinfo {pages} {B571} (\bibinfo {year} {1964})}\BibitemShut {NoStop}%
\bibitem [{\citenamefont {Telkamp}(2016)}]{TelkampPhysRevD.94.043520}%
  \BibitemOpen
  \bibfield  {author} {\bibinfo {author} {\bibfnamefont {H.}~\bibnamefont
  {Telkamp}},\ }\href@noop {} {\bibfield  {journal} {\bibinfo  {journal} {Phys.
  Rev. D}\ }\textbf {\bibinfo {volume} {94}},\ \bibinfo {pages} {043520}
  (\bibinfo {year} {2016})}\BibitemShut {NoStop}%
\bibitem [{\citenamefont {Eddington}(1920)}]{Eddington}%
  \BibitemOpen
  \bibfield  {author} {\bibinfo {author} {\bibfnamefont {A.}~\bibnamefont
  {Eddington}},\ }\href@noop {} {\emph {\bibinfo {title} {Space, Time, and
  Gravitation}}}\ (\bibinfo  {publisher} {London: Cambridge University Press},\
  \bibinfo {year} {1920})\BibitemShut {NoStop}%
\bibitem [{\citenamefont {Fernflores}(2012)}]{StanfEncyclEquivalencePrinciple}%
  \BibitemOpen
  \bibfield  {author} {\bibinfo {author} {\bibfnamefont {F.}~\bibnamefont
  {Fernflores}},\ }in\ \href@noop {} {\emph {\bibinfo {booktitle} {The Stanford
  Encyclopedia of Philosophy}}},\ \bibinfo {editor} {edited by\ \bibinfo
  {editor} {\bibfnamefont {E.~N.}\ \bibnamefont {Zalta}}}\ (\bibinfo
  {publisher} {Metaphysics Research Lab, Stanford University},\ \bibinfo {year}
  {2012})\ \bibinfo {edition} {spring 2012}\ ed.\BibitemShut {Stop}%
\bibitem [{\citenamefont {Ellis}\ and\ \citenamefont
  {Uzan}(2015)}]{Ellis:2015wdi}%
  \BibitemOpen
  \bibfield  {author} {\bibinfo {author} {\bibfnamefont {G.~F.~R.}\
  \bibnamefont {Ellis}}\ and\ \bibinfo {author} {\bibfnamefont {J.-P.}\
  \bibnamefont {Uzan}},\ }\href {\doibase 10.1016/j.crhy.2015.07.005}
  {\bibfield  {journal} {\bibinfo  {journal} {Comptes Rendus Physique}\
  }\textbf {\bibinfo {volume} {16}},\ \bibinfo {pages} {928} (\bibinfo {year}
  {2015})},\ \Eprint {http://arxiv.org/abs/1612.01084} {arXiv:1612.01084
  [gr-qc]} \BibitemShut {NoStop}%
\bibitem [{\citenamefont {Chodorowski}(2011)}]{Chodorowski:2009cb}%
  \BibitemOpen
  \bibfield  {author} {\bibinfo {author} {\bibfnamefont {M.}~\bibnamefont
  {Chodorowski}},\ }\href {\doibase 10.1111/j.1365-2966.2010.18154.x}
  {\bibfield  {journal} {\bibinfo  {journal} {Mon. Not. Roy. Astron. Soc.}\
  }\textbf {\bibinfo {volume} {413}},\ \bibinfo {pages} {585} (\bibinfo {year}
  {2011})},\ \Eprint {http://arxiv.org/abs/0911.3536} {arXiv:0911.3536
  [astro-ph.CO]} \BibitemShut {NoStop}%
\bibitem [{\citenamefont {Faraoni}(2010)}]{Faraoni:2009buREDSHIFT}%
  \BibitemOpen
  \bibfield  {author} {\bibinfo {author} {\bibfnamefont {V.}~\bibnamefont
  {Faraoni}},\ }\href {\doibase 10.1007/s10714-009-0885-8} {\bibfield
  {journal} {\bibinfo  {journal} {Gen. Rel. Grav.}\ }\textbf {\bibinfo {volume}
  {42}},\ \bibinfo {pages} {851} (\bibinfo {year} {2010})},\ \Eprint
  {http://arxiv.org/abs/0908.3431} {arXiv:0908.3431 [gr-qc]} \BibitemShut
  {NoStop}%
\bibitem [{\citenamefont {Melia}(2012)}]{Melia:2012ic}%
  \BibitemOpen
  \bibfield  {author} {\bibinfo {author} {\bibfnamefont {F.}~\bibnamefont
  {Melia}},\ }\href {\doibase 10.1111/j.1365-2966.2012.20714.x} {\bibfield
  {journal} {\bibinfo  {journal} {Mon. Not. Roy. Astron. Soc.}\ }\textbf
  {\bibinfo {volume} {422}},\ \bibinfo {pages} {1418} (\bibinfo {year}
  {2012})},\ \Eprint {http://arxiv.org/abs/1202.0775} {arXiv:1202.0775
  [astro-ph.CO]} \BibitemShut {NoStop}%
\bibitem [{\citenamefont {Szabados}(2009)}]{Szabados2009}%
  \BibitemOpen
  \bibfield  {author} {\bibinfo {author} {\bibfnamefont {L.~B.}\ \bibnamefont
  {Szabados}},\ }\href {\doibase 10.12942/lrr-2009-4} {\bibfield  {journal}
  {\bibinfo  {journal} {Living Reviews in Relativity}\ }\textbf {\bibinfo
  {volume} {12}} (\bibinfo {year} {2009}),\ 10.12942/lrr-2009-4}\BibitemShut
  {NoStop}%
\bibitem [{\citenamefont {{Rindler}}(2006)}]{Rindler}%
  \BibitemOpen
  \bibfield  {author} {\bibinfo {author} {\bibfnamefont {W.}~\bibnamefont
  {{Rindler}}},\ }\href@noop {} {\emph {\bibinfo {title} {{Relativity}}}}\
  (\bibinfo  {publisher} {Oxford: University Press},\ \bibinfo {year}
  {2006})\BibitemShut {NoStop}%
\bibitem [{\citenamefont {{Faraoni}}(2011)}]{FaraoniPhysRevD.84.024003}%
  \BibitemOpen
  \bibfield  {author} {\bibinfo {author} {\bibfnamefont {V.}~\bibnamefont
  {{Faraoni}}},\ }\href@noop {} {\bibfield  {journal} {\bibinfo  {journal}
  {Phys. Rev. D}\ }\textbf {\bibinfo {volume} {84}},\ \bibinfo {pages} {024003}
  (\bibinfo {year} {2011})}\BibitemShut {NoStop}%
\bibitem [{\citenamefont {Binetruy}\ and\ \citenamefont
  {Helou}(2015)}]{Binetruy:2014ela}%
  \BibitemOpen
  \bibfield  {author} {\bibinfo {author} {\bibfnamefont {P.}~\bibnamefont
  {Binetruy}}\ and\ \bibinfo {author} {\bibfnamefont {A.}~\bibnamefont
  {Helou}},\ }\href {\doibase 10.1088/0264-9381/32/20/205006} {\bibfield
  {journal} {\bibinfo  {journal} {Class. Quant. Grav.}\ }\textbf {\bibinfo
  {volume} {32}},\ \bibinfo {pages} {205006} (\bibinfo {year} {2015})},\
  \Eprint {http://arxiv.org/abs/1406.1658} {arXiv:1406.1658 [gr-qc]}
  \BibitemShut {NoStop}%
\bibitem [{\citenamefont {Mach}(1893)}]{Mach}%
  \BibitemOpen
  \bibfield  {author} {\bibinfo {author} {\bibfnamefont {E.}~\bibnamefont
  {Mach}},\ }\href@noop {} {\emph {\bibinfo {title} {{The Science of Mechanics:
  A Critical Historical Account of its Development}}}}\ (\bibinfo  {publisher}
  {Open Court, La Salle, IL},\ \bibinfo {year} {1893})\BibitemShut {NoStop}%
\bibitem [{\citenamefont {Huggett}\ and\ \citenamefont
  {Hoefer}(2017)}]{sep-RelationalTheories}%
  \BibitemOpen
  \bibfield  {author} {\bibinfo {author} {\bibfnamefont {N.}~\bibnamefont
  {Huggett}}\ and\ \bibinfo {author} {\bibfnamefont {C.}~\bibnamefont
  {Hoefer}},\ }in\ \href@noop {} {\emph {\bibinfo {booktitle} {The Stanford
  Encyclopedia of Philosophy}}},\ \bibinfo {editor} {edited by\ \bibinfo
  {editor} {\bibfnamefont {E.~N.}\ \bibnamefont {Zalta}}}\ (\bibinfo
  {publisher} {Metaphysics Research Lab, Stanford University},\ \bibinfo {year}
  {2017})\ \bibinfo {edition} {spring 2017}\ ed.\BibitemShut {Stop}%
\bibitem [{\citenamefont {{Ann M. Hentschel
  (translation)}}(1998)}]{CollectedPapersEinsteinVol8}%
  \BibitemOpen
  \bibfield  {author} {\bibinfo {author} {\bibnamefont {{Ann M. Hentschel
  (translation)}}},\ }\href@noop {} {\emph {\bibinfo {title} {{The Collected
  Papers of Albert Einstein, Volume 8 (English):The Berlin Years:
  Correspondence, 1914-1918. }}}}\ (\bibinfo  {publisher} {Princeton University
  Press},\ \bibinfo {year} {1998})\BibitemShut {NoStop}%
\bibitem [{\citenamefont {Sciama}(1953)}]{Sciama}%
  \BibitemOpen
  \bibfield  {author} {\bibinfo {author} {\bibfnamefont {D.~W.}\ \bibnamefont
  {Sciama}},\ }\href@noop {} {\bibfield  {journal} {\bibinfo  {journal}
  {MNRAS}\ }\textbf {\bibinfo {volume} {113}},\ \bibinfo {pages} {34} (\bibinfo
  {year} {1953})}\BibitemShut {NoStop}%
\bibitem [{\citenamefont {Berkeley}(1992)}]{berkeley}%
  \BibitemOpen
  \bibfield  {author} {\bibinfo {author} {\bibfnamefont {G.}~\bibnamefont
  {Berkeley}},\ }\href@noop {} {\emph {\bibinfo {title} {{De Motu and The
  Analyst}}}}\ (\bibinfo  {publisher} {Springer/Kluwer Academic Publishers},\
  \bibinfo {year} {1992})\BibitemShut {NoStop}%
\bibitem [{\citenamefont {Brans}\ and\ \citenamefont
  {Dicke}(1961)}]{BransDicke}%
  \BibitemOpen
  \bibfield  {author} {\bibinfo {author} {\bibfnamefont {C.}~\bibnamefont
  {Brans}}\ and\ \bibinfo {author} {\bibfnamefont {R.~H.}\ \bibnamefont
  {Dicke}},\ }\href {\doibase 10.1103/PhysRev.124.925} {\bibfield  {journal}
  {\bibinfo  {journal} {Phys. Rev.}\ }\textbf {\bibinfo {volume} {124}},\
  \bibinfo {pages} {925} (\bibinfo {year} {1961})}\BibitemShut {NoStop}%
\bibitem [{\citenamefont {{Einstein}}(1916)}]{Einstein1916}%
  \BibitemOpen
  \bibfield  {author} {\bibinfo {author} {\bibfnamefont {A.}~\bibnamefont
  {{Einstein}}},\ }\href {\doibase 10.1002/andp.19163540702} {\bibfield
  {journal} {\bibinfo  {journal} {Annalen der Physik}\ }\textbf {\bibinfo
  {volume} {354}},\ \bibinfo {pages} {769} (\bibinfo {year}
  {1916})}\BibitemShut {NoStop}%
\bibitem [{\citenamefont {Einstein}(1918)}]{einstein1918prinzipielles}%
  \BibitemOpen
  \bibfield  {author} {\bibinfo {author} {\bibfnamefont {A.}~\bibnamefont
  {Einstein}},\ }\href@noop {} {\bibfield  {journal} {\bibinfo  {journal}
  {Annalen der Physik}\ }\textbf {\bibinfo {volume} {360}},\ \bibinfo {pages}
  {241} (\bibinfo {year} {1918})}\BibitemShut {NoStop}%
\bibitem [{\citenamefont {{Schroedinger}}(1925)}]{Schroedinger}%
  \BibitemOpen
  \bibfield  {author} {\bibinfo {author} {\bibfnamefont {E.}~\bibnamefont
  {{Schroedinger}}},\ }\href@noop {} {\bibfield  {journal} {\bibinfo  {journal}
  {Annalen der Physik}\ }\textbf {\bibinfo {volume} {382}},\ \bibinfo {pages}
  {325} (\bibinfo {year} {1925})}\BibitemShut {NoStop}%
\bibitem [{\citenamefont {{Barbour}}\ and\ \citenamefont
  {{Pfister}}(1995)}]{Barbour}%
  \BibitemOpen
  \bibinfo {editor} {\bibfnamefont {J.~B.}\ \bibnamefont {{Barbour}}}\ and\
  \bibinfo {editor} {\bibfnamefont {H.}~\bibnamefont {{Pfister}}},\ eds.,\
  \href@noop {} {\emph {\bibinfo {title} {{Mach's Principle: From Newton's
  Bucket to Quantum Gravity.}}}}\ (\bibinfo  {publisher} {Birkenhauser,
  Boston},\ \bibinfo {year} {1995})\BibitemShut {NoStop}%
\bibitem [{\citenamefont {Parikh}(2002)}]{PARIKH2002189}%
  \BibitemOpen
  \bibfield  {author} {\bibinfo {author} {\bibfnamefont {M.~K.}\ \bibnamefont
  {Parikh}},\ }\href {\doibase http://dx.doi.org/10.1016/S0370-2693(02)02701-6}
  {\bibfield  {journal} {\bibinfo  {journal} {Physics Letters B}\ }\textbf
  {\bibinfo {volume} {546}},\ \bibinfo {pages} {189 } (\bibinfo {year}
  {2002})}\BibitemShut {NoStop}%
\bibitem [{\citenamefont {Bondi}\ and\ \citenamefont {Gold}(1948)}]{BondiGold}%
  \BibitemOpen
  \bibfield  {author} {\bibinfo {author} {\bibfnamefont {H.}~\bibnamefont
  {Bondi}}\ and\ \bibinfo {author} {\bibfnamefont {T.}~\bibnamefont {Gold}},\
  }\href {\doibase "10.1093/mnras/108.3.252"} {\bibfield  {journal} {\bibinfo
  {journal} {Monthly Notices of the Royal Astronomical Society}\ }\textbf
  {\bibinfo {volume} {108}},\ \bibinfo {pages} {252} (\bibinfo {year}
  {1948})}\BibitemShut {NoStop}%
\bibitem [{\citenamefont {Hoyle}(1948)}]{Hoyle}%
  \BibitemOpen
  \bibfield  {author} {\bibinfo {author} {\bibfnamefont {F.}~\bibnamefont
  {Hoyle}},\ }\href {\doibase "10.1093/mnras/108.5.372"} {\bibfield  {journal}
  {\bibinfo  {journal} {Monthly Notices of the Royal Astronomical Society}\
  }\textbf {\bibinfo {volume} {108}},\ \bibinfo {pages} {372} (\bibinfo {year}
  {1948})}\BibitemShut {NoStop}%
\bibitem [{\citenamefont {Kolb}(1989)}]{kolb1989coasting}%
  \BibitemOpen
  \bibfield  {author} {\bibinfo {author} {\bibfnamefont {E.~W.}\ \bibnamefont
  {Kolb}},\ }\href@noop {} {\bibfield  {journal} {\bibinfo  {journal} {The
  Astrophysical Journal}\ }\textbf {\bibinfo {volume} {344}},\ \bibinfo {pages}
  {543} (\bibinfo {year} {1989})}\BibitemShut {NoStop}%
\bibitem [{\citenamefont {Layzer}(1954)}]{layzer}%
  \BibitemOpen
  \bibfield  {author} {\bibinfo {author} {\bibfnamefont {D.}~\bibnamefont
  {Layzer}},\ }\href@noop {} {\bibfield  {journal} {\bibinfo  {journal}
  {Astronomical Journal}\ }\textbf {\bibinfo {volume} {59}},\ \bibinfo {pages}
  {268} (\bibinfo {year} {1954})}\BibitemShut {NoStop}%
\bibitem [{\citenamefont {Ade}\ \emph {et~al.}(2016)\citenamefont {Ade} \emph
  {et~al.}}]{Planck2015}%
  \BibitemOpen
  \bibfield  {author} {\bibinfo {author} {\bibfnamefont {P.~A.~R.}\
  \bibnamefont {Ade}} \emph {et~al.} (\bibinfo {collaboration} {Planck}),\
  }\href {\doibase 10.1051/0004-6361/201525830} {\bibfield  {journal} {\bibinfo
   {journal} {Astron. Astrophys.}\ }\textbf {\bibinfo {volume} {594}},\
  \bibinfo {pages} {A13} (\bibinfo {year} {2016})},\ \Eprint
  {http://arxiv.org/abs/1502.01589} {arXiv:1502.01589 [astro-ph.CO]}
  \BibitemShut {NoStop}%
\bibitem [{\citenamefont {{Suzuki}}\ \emph {et~al.}(2012)\citenamefont
  {{Suzuki}}, \citenamefont {{Lidman}} \emph
  {et~al.}}]{SupernovaCosmologyProject2012ApJ}%
  \BibitemOpen
  \bibfield  {author} {\bibinfo {author} {\bibfnamefont {D.}~\bibnamefont
  {{Suzuki}}, \bibfnamefont {N.~{Rubin}}}, \bibinfo {author} {\bibfnamefont
  {C.}~\bibnamefont {{Lidman}}},  \emph {et~al.},\ }\href {\doibase
  10.1088/0004-637X/746/1/85} {\bibfield  {journal} {\bibinfo  {journal}
  {\apj}\ }\textbf {\bibinfo {volume} {746}},\ \bibinfo {eid} {85} (\bibinfo
  {year} {2012})},\ \Eprint {http://arxiv.org/abs/1105.3470} {arXiv:1105.3470
  [astro-ph.CO]} \BibitemShut {NoStop}%
\bibitem [{\citenamefont {Wagenmakers}\ and\ \citenamefont
  {Farrell}(2004)}]{Wagenmakers2004}%
  \BibitemOpen
  \bibfield  {author} {\bibinfo {author} {\bibfnamefont {E.-J.}\ \bibnamefont
  {Wagenmakers}}\ and\ \bibinfo {author} {\bibfnamefont {S.}~\bibnamefont
  {Farrell}},\ }\href {\doibase 10.3758/BF03206482} {\bibfield  {journal}
  {\bibinfo  {journal} {Psychonomic Bulletin {\&} Review}\ }\textbf {\bibinfo
  {volume} {11}},\ \bibinfo {pages} {192} (\bibinfo {year} {2004})}\BibitemShut
  {NoStop}%
\bibitem [{\citenamefont {{Harrison}}(1981)}]{Harrison}%
  \BibitemOpen
  \bibfield  {author} {\bibinfo {author} {\bibfnamefont {E.~R.}\ \bibnamefont
  {{Harrison}}},\ }\href@noop {} {\emph {\bibinfo {title} {{Cosmology. The
  science of the universe}}}}\ (\bibinfo  {publisher} {Cambridge: University
  Press},\ \bibinfo {year} {1981})\BibitemShut {NoStop}%
\end{thebibliography}

%

\end{document}